# Training Data Set Refinement for the Machine Learning Potential of Li–Si Alloys via Structural Similarity Analysis

*Nan Xu,*[a] *Chen Li,*[a] *Mandi Fang,*[a] *Qing Shao,*[b] *Yao Shi,*[a] *Yingying Lu,*[a] *Yi He,* [a,c*]

a College of Chemical and Biological Engineering, Institute of Zhejiang University-Quzhou, Key Laboratory of Biomass Chemical Engineering of Ministry of Education, Zhejiang University, Hangzhou 310027, China

b Chemical and Materials Engineering Department, University of Kentucky, Lexington, KY 40506, USA

c Department of Chemical Engineering, University of Washington, Seattle, WA 98195, USA

ABSTRACT

Machine learning potential enables molecular dynamics simulations of systems beyond the capability of classical force fields. The traditional approach to develop structural sets for training machine learning potential typically generate a great number of redundant configurations, which will result in unnecessary computational costs. This work investigates the possibility of reducing redundancy in an initial data set containing 6183 configurations for a Li–Si machine learning potential. Starting from the initial data set, we constructed a series of subsets ranging from 25 to 1500 configurations by combining a structural similarity analysis algorithm and the farthest point sampling method. Results show that the machine learning potential trained from a data set containing 400 configurations can achieve an accuracy comparable to the one developed from the initial data set of 6183 configurations in describing potential energies, atomic forces, and structural properties of Li–Si systems. In addition, the redundancy reducing approach also demonstrates advantages over the classic stochastic method for constructing a concise training data set for Li–Si systems.

Keywords: Li–Si, machine learning potential, structural similarity, data set refinement

## 1. Introduction

Machine learning (ML) potentials are gaining lots of attractions in recent years as they demonstrate capabilities which are otherwise unattainable with classical force fields (FFs) for molecular dynamics (MD) simulations.[1-4] While classical FFs such as the OPLS-AA,[5-6] Charmm,[7-8] and Amber[9-10] have achieved great success in describing organic and biological systems, these FFs typically have limitations which are originated from the preset functions[11-13] used to represent the potential energy, for instance, it is not feasible to use harmonic functions to describe simulation systems involving chemical reactions. The reactive force field (ReaxFF)[14-15] provides bond formation/breaking features based on the classical FFs by introducing new function forms. However, developing ReaxFF is often a time-consuming process and requires well-trained professionals.[16] Quantum mechanical (QM) calculations can describe atomic interactions in great precision. However, the massive computational demand of QM hinders the applications in investigating dynamic properties of large systems. ML potentials utilize a combination of molecular descriptors and ML models to build connections between structures and their properties.[17] A well-developed ML potential can not only be comparable to classical FFs in speed but also be at the same level as QM calculations in accuracy. Recently, ML potentials have shown their suitability for complex cases such as reactive or metallic systems.[18-22]

One challenge in developing a high-accurate ML potential is to construct structural sets with low sample redundancy.[23-24] The traditional approach to construct data set such as the *ab initio* molecular dynamics (AIMD)[25-26] inevitably generates redundant configurations. These redundant configurations may lead to overfitting issues of ML potentials.[24] In addition, they will cause substantial needless computational cost as their properties such as the potential energies and atomic forces[27] require compute-intensive density functional theory (DFT) calculations.[28-29] Therefore, reducing redundancy in an initial data set before DFT calculations would significantly lower the computational demands while maintaining the accuracy of the ML potentials.

Some efforts have been made to reduce the redundancy in training data sets.[30-32] Stochastic sampling is a classic approach for this purpose.[33-34] However, it lacks of reliable capability of removing overrepresented samples from the initial data set,[32] as some essential and representative configurations may be missing due to the randomness.

Physics-based sampling from the initial data set is considered to be an effective method for reducing redundancy.[17, 32, 35] Michaelides and coworkers[35] proposed a method to construct a concise data set for hexagonal boron nitride (hBN) by selecting representative configurations from an initial data set. They mapped 22000 hBN configurations including monolayers, bilayers, nanotubes and bulks from AIMD simulations into a similarity space[30], where the similarities between configurations were computed by the smooth overlap of atomic positions (SOAP) kernels.[4, 27, 36] 2000 configurations were then selected by using a combination of the farthest point sampling (FPS)[37-38] and manual structure selection[35]. They found the ML potential developed from this reduced data set predicts similar formation energies, and mechanical properties as those calculated from DFT simulations.

In our previous work, we trained a Deep Potential (DP) for Li–Si systems with a data set constructed from MD trajectories and atom distortions.[39] The DP can predict potential energies and structural properties of Li–Si systems with an accuracy similar to the DFT calculations. The training data set for developing this potential contains more than 6000 samples, which poses a significant computational burden when calculating various structure-related properties with DFT. Therefore, it is of great significance to refine the training data set without causing any detrimental effects on the reliability of the corresponding ML potentials. Herein, we refined the training data set used in our previous work by constructing different subsets from the data set using a combination of the structural similarity algorithm and the FPS method, and then trained ML potentials from the subsets using the framework of the moment tensor potential (MTP).[3, 40] We investigated how the size of the subsets affect the redundancy reducing of the initial data set, and evaluate the accuracy of the MTPs trained with small subsets such as containing 200 and 400 configurations in predicting potential energies, atomic forces, and radial distribution functions (RDF) for different Li–Si alloys. The rest of the paper is organized as follows. Section 2 describes the details of MTPs, structural similarity analysis algorithm, FPS method, and QM calculations. Section 3 exhibits the results and discussion, and Section 4 presents the conclusion.

## 2. Simulation Details

### 2.1 Development of moment tensor potentials

The ML potential for Li–Si alloys in this work was developed under the framework of the MTP proposed by Shapeev and coworkers.[3, 40] This potential was reported to have reasonable precisions even using small training data sets of 100–200 structures. [27] The MTP calculates the potential energy by summing energy contribution ($V_i$) from atom $i$ in a system of $n$ atoms, which is also adopted in other ML potentials.[2, 4, 41] $V_i$ is then linearly expanded through a set of basis functions:

$$V_i(\boldsymbol{n}_i) = \sum_{\alpha} \xi_\alpha B_{\boldsymbol{\alpha}}(\boldsymbol{n}_i) \tag{1}$$

where $\xi_\alpha$ is parameters to be determined in the training process, $\boldsymbol{n}_i$ is a set of neighboring atoms of $i$, and $B_{\boldsymbol{\alpha}}(\boldsymbol{n}_i)$ is a basis function. $B_{\boldsymbol{\alpha}}(\boldsymbol{n}_i)$ is constructed using a combination of a series of moment tensor descriptors $M_{\mu,\nu}(\boldsymbol{n}_i)$[3, 40] which contain radial and angular information between atom $i$ and its neighboring atoms within $\boldsymbol{n}_i$. An accuracy level ($\mathrm{lev}_{\mathrm{max}}$) of 16 was used in this work, which determines the size of basis functions by specifying possible values of $\mu$ and $\nu$. In addition, a cutoff of 5 Å, a minimal distance of 2 Å, and a radial basis size of 6 were used in the construction of MTP. A maximum iteration epoch of 1000 was set in the training.

### 2.2 Construction of training, validation and testing data sets

The training, validation, and testing data sets in this work are taken from those used for DP models of Li–Si alloys in our previous work.[39] The raw training data for DP models contains 6258 Li–Si configurations. 5250 configurations of them were from melting-quenching MD simulations of crystalline $Li_1Si_{64}$, $Li_1Si_3$, $Li_1Si_1$, $Li_{13}Si_4$ and $Li_{54}Si_1$, while 600 configurations were generated by displacing atoms and reshaping boxes of crystalline $Li_1Si_1$. In addition, 408 configurations were explored by performing five iterations of active-learning strategy.[42] A looser force error criterion ($\delta_{\mathrm{Fmax}}$) of 0.4 $\mathrm{eV/Å}$ was used in the fifth iteration while compared to that of 0.25 $\mathrm{eV/Å}$ used in the first four iterations.[39] In this work, we used 0.25 $\mathrm{eV/Å}$ as the force error criterion for the five active-learning iterations, and 75 out of 408 configurations were ruled out. Therefore, we obtained a data set containing 6183 Li–Si configurations which is referred as the initial data set. The number of configurations for $Li_1Si_{64}$, $Li_1Si_3$, $Li_1Si_1$, $Li_{13}Si_4$ and $Li_{54}Si_1$ are 580, 572, 3926, 580 and 525, respectively, where $Li_1Si_1$ accounts for 63.5% of the initial data set and other compositions share similar percentages of about 9.0%. Structure-related properties such as the potential energies and atomic forces for each configuration remain the same as those in the previous work. The initial data set was divided into two parts: the training data set of 3092 configurations, and the validation data set of 3091 configurations.

The potential energy deviation $\Delta E$ and atomic force deviation $\Delta F$ for training and validation are calculated as follows:

$$\Delta E = \sqrt{\frac{1}{K}\sum_{k=1}^{K}\left(\frac{E_{\mathrm{MTP}}^{k} - E_{\mathrm{DFT}}^{k}}{N_k}\right)^2} \quad (2)$$

$$\Delta F = \sqrt{\frac{1}{K}\sum_{k=1}^{K}\frac{1}{3N_k}\sum_{i=1}^{N_k}\left|\mathbf{f}_{\mathrm{MTP}}^{i} - \mathbf{f}_{\mathrm{DFT}}^{i}\right|^2} \quad (3)$$

where $K$ is the number of configurations, $N_k$ is the number of atoms in the $k^{th}$ configuration. $E_{\mathrm{MTP}}^{k}$ and $E_{\mathrm{DFT}}^{k}$ are the potential energies of the $k^{th}$ configuration predicted by MTP and calculated by DFT, while $\mathbf{f}_{\mathrm{MTP}}^{i}$ and $\mathbf{f}_{\mathrm{DFT}}^{i}$ are the forces on atom $i$ predicted by MTP and calculated by DFT, respectively.

The testing data set from the previous work was also used to evaluate the generalization ability of the trained MTPs.[39] This data set contains amorphous Si, $Li_1Si_1$, $Li_2Si_1$, and $Li_7Si_2$ with 501 configurations for each composition. The Euclidean norm of atomic forces on Li or Si atoms predicted by MTP, defined as $||\mathbf{f}_{\mathrm{MTP}}||$, were compared with those calculated by DFT ($||\mathbf{f}_{\mathrm{DFT}}||$).

**2.3 Structural similarity calculation**

The similarity between two given configurations in the training data set was computed by averaging all similarities between local environment $\chi$ in configuration A and $\chi'$ in configuration B. $\chi$ refers to a central atom $i$ and neighboring atoms within a cutoff $r_{\mathrm{c}}$. In this work, $r_{\mathrm{c}}$ was set to be 3 Å, and a transition width of 0.5 Å was augmented. This cutoff is larger than the distance of the first peak in Si–Si, Li–Si and Li–Li radial distribution functions (RDF) profiles[39], and therefore, contains the coordination information for both Li and Si atoms. The SOAP kernels were used to calculate the similarity between $\chi$ and $\chi'$ by integrating the overlap of the local neighbor densities in the two environments.[4, 30, 36] The local density of a central atom $i$ is described as a sum of Gaussian functions centered over all neighboring atoms,

$$\rho_i(\boldsymbol{r}) = \sum_j \exp\left(-\frac{\left|\boldsymbol{r}-\boldsymbol{r}_{ij}\right|^2}{2\sigma^2}\right) \tag{4}$$

where $\boldsymbol{r}_{ij}$ is the distance vector between atom $i$ and $j$ within $r_{\mathrm{c}}$ of $i$, and σ is a smearing parameter with a value of 0.4.

The SOAP kernel $\tilde{k}$ is obtained by integrating the overlap of $\rho_{\chi}$ and $\rho_{\chi'}$,

$$\tilde{k}(\chi,\chi') = \int d\widehat{\boldsymbol{R}} \left| \int \rho_{\chi}(\boldsymbol{r})\rho_{\chi'}\left(\widehat{\boldsymbol{R}}\boldsymbol{r}\right) d\boldsymbol{r} \right|^2 \tag{5}$$

where $\widehat{\boldsymbol{R}}$ is a set of rotations vectors for the purpose of rotationally invariance.[30] The final SOAP kernel is normalized as follows,

$$k(\chi,\chi') = \frac{\tilde{k}(\chi,\chi')}{\sqrt{\tilde{k}(\chi,\chi)\tilde{k}(\chi',\chi')}} \tag{6}$$

Averaging $k(\chi,\chi')$ kernels for all $\chi$–$\chi'$ pairs give the overall similarity between A and B,

$$K(A,B) = \frac{1}{n_A n_B}\sum_{i}^{n_A}\sum_{j}^{n_B} k(\chi_i,\chi'_j) \tag{7}$$

where $n_A$ and $n_B$ are the number of local environments in A and B.

**2.4 Similarity diagram of configurations**

Similarity diagrams are used to intuitively present the distributions of configurations from the training data set or a specific FPS subset. The location of a configuration in a similarity diagram is determined by the similarities respect to the configurations of $Li_1Si_{64}$ and $Li_{54}Si_1$. More specifically, the x coordinate of a configuration is calculated by averaging its similarities with respect to each configuration with a composition of $Li_1Si_{64}$, while the y coordinate is calculated by averaging its similarity with respect to each configuration with a composition of $Li_{54}Si_1$.

**2.5 Scheme of the FPS method**

The FPS method uses distances between samples for data selections. In this work, the distance between configurations A and B is calculated according to their similarity $K(A,B)$,[30]

$$d = \sqrt{2 - 2 * K(A,B)} \tag{8}$$

The FPS method typically starts with a random or manually selected point as the first sample, and iteratively selects the farthest point from the already selected samples.[43] Take the construction of a subset containing N configurations for example, the FPS method chooses a configuration which is closest to the training data set as the first point, where the distance is defined as:

$$\bar{d}_i = \frac{\sum_{j\neq i}^{N} d_{ij}}{N-1} \tag{9}$$

Then, the FPS method iteratively selects the remaining N-1 configurations. In each step, a configuration which is farthest away from all configurations in the subset is selected and added into the current subset. The distance between configuration $i$ and all configurations in a subset is defined as:

$$d_i^{min} = \min_j d_{i,j} \tag{10}$$

where $j$ runs through all configurations in the current subset. **Figure 1** shows the scheme of selecting 3 data from a data set of 4 configurations according to their pairwise similarities using the FPS method.

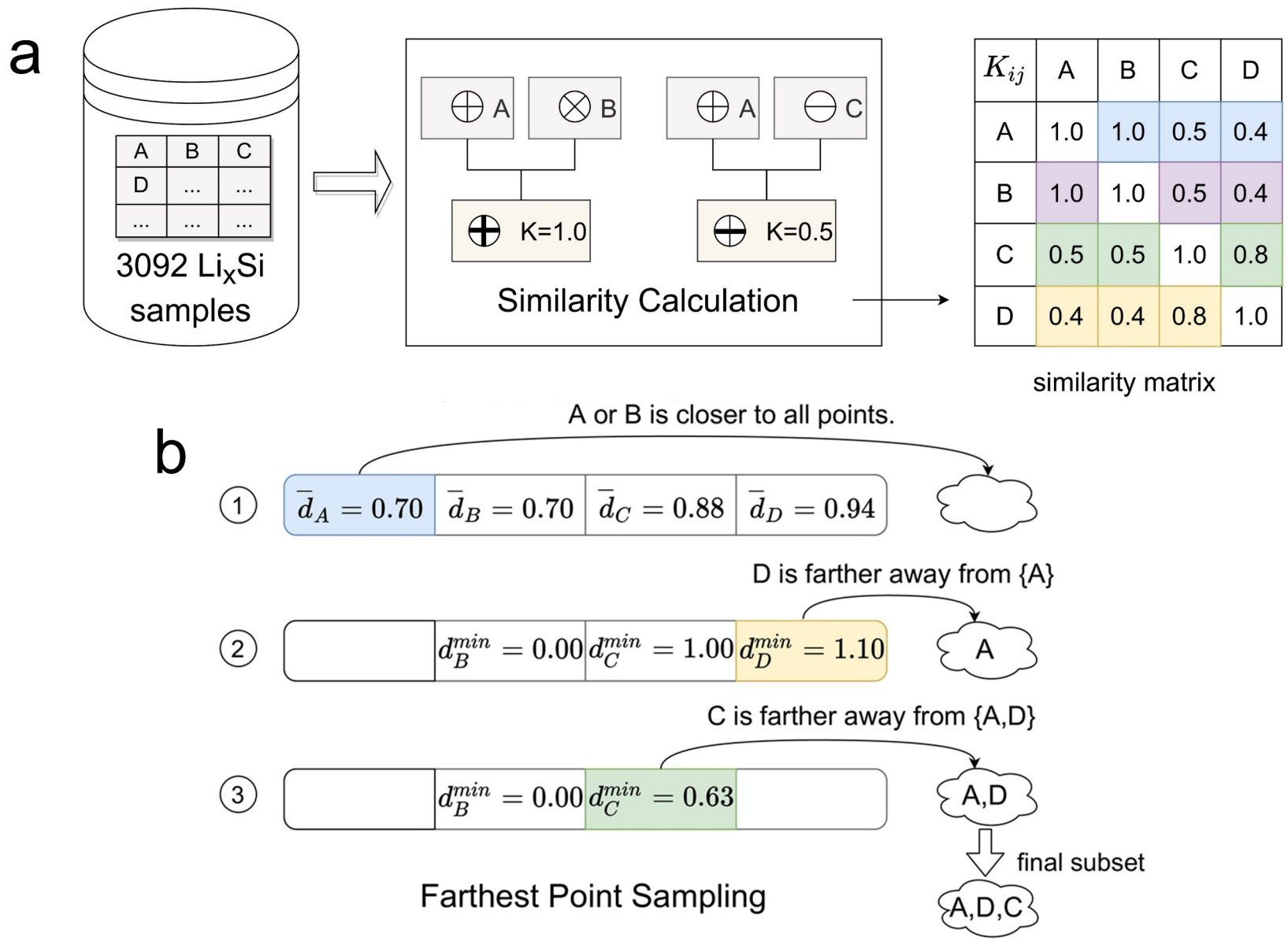

**Figure 1.** (a) Flowchart to calculate pairwise structural similarities for a subset of 4 configurations. ⊕ and ⊗ stand for two rotational invariant configurations, while ⊕ and ⊖ stand for two different configurations. The similarity value[30] between A and B

is calculated to be 1.0. (b) Scheme of selecting 3 data from a data set of 4 configurations according to the pairwise similarities using the FPS method.

### 2.6 Construction of subsets

Several subsets ranging from 25 to 1500 configurations were constructed from the training data set using a combination of the structural similarity analysis and the FPS method, respectively. Only the subsets containing 200 and 400 configurations were used for developing new MTPs. In addition, 3 subsets containing 400 configurations were generated from the training data set by the stochastic sampling method, and then used for training other three MTPs. The subsets constructed from the training data set by the FPS method are referred as the FPS subsets, while the subsets generated by the stochastic sampling method are referred as the SS subsets.

### 2.7 Structural representativeness of subsets

Ideally, all representative configurations in the initial training set should be retained when constructing a subset for the training of machine learning potentials. Therefore, we defined the representativeness of a subset to measure the extend to which the structural features in the initial data set are included in the subset as follow:

$$\text{REP} = \frac{N_{rep}}{N} \times 100\% \tag{11}$$

where $N$ is the total number of configurations in the training data set, $N_{rep}$ is the number of configurations in the training data set which have been represented by a configuration in the subset. A configuration in the training data set will be marked as represented, if any of the configurations in the subset is at a distance of less than 0.1. A larger REP indicates a better subset in representing the training set.

### 2.8 MTP-MD simulations

MD simulations with MTPs (MTP-MD) were performed for amorphous Si, $Li_1Si_1$, $Li_2Si_1$, and $Li_7Si_2$ at 298 K. The NVT ensemble was used in the MTP-MD simulations with the Nose−Hoover thermostat algorisms.[44] The production runs have a duration of 500 ps after a 100-ps equilibrium period with a time step of 0.5 fs. The large-scale atomic/molecular massively parallel simulator (LAMMPS)[45] and the MTP framework[3, 40] were used for the MTP-MD calculations.

### 2.9 AIMD simulations

AIMD simulations were also performed using the same initial configurations as the MTP-MD simulations. The exchange-correlation potential uses the Perdew−Burke−Ernzerhof (PBE) functionals.[46] The projector augmented wave (PAW) approach by Blöchl[47] was used for the core electrons with a kinetic energy cutoff of 520 eV. Configurations of $1s^1 2s^1 2p^1$ and $3s^2 3p^2$ are used for the valence-electrons of Li and Si, respectively. The Γ point was used to sample the Brillouin zone for AIMD. The Gaussian smearing with a parameter σ of 0.05 eV was used for electronic iterations and the energy convergence criterion was set to $1.0 \times 10^{-7}$ eV. In addition, the zero-damping corrections were used for the long-range dispersion.[48] The production runs have a duration of 80 ps after a 20 ps equilibrium period with a time step of 1.0 fs. The AIMD simulation were also performed in the NVT ensemble at 298 K using the Nose−Hoover thermostat algorisms.[44] The Vienna ab initio simulation package (VASP) [49-50] of version 5.4.4 was used in the AIMD simulations.

### 2.10 Comparison of RDFs

The RDFs of Li–Li, Li–Si and Si–Si pairs were calculated for amorphous Si, $Li_1Si_1$, $Li_2Si_1$, and $Li_7Si_2$ at 298 K in MTP-MD simulations, and the position ($r$) and the height ($h$) of the first RDF peak were compared with those computed from AIMD simulations.

$\Delta_r$ and $\delta_h$ show the position deviations and the relative height deviations of the first RDF peaks between MTP-MD and AIMD simulations, which are defined as follows:

$$\Delta_r = r_{\mathrm{MTP-MD}} - r_{\mathrm{AIMD}} \tag{12}$$

$$\delta_h = \frac{h_{\mathrm{MTP-MD}} - h_{\mathrm{AIMD}}}{h_{\mathrm{AIMD}}} \tag{13}$$

where $r_{\mathrm{MTP-MD}}$ and $h_{\mathrm{MTP-MD}}$ are the position and height of the first RDF peaks in MTP-MD simulations, while $r_{\mathrm{AIMD}}$ and $h_{\mathrm{AIMD}}$ are the position and height of the first RDF peaks in AIMD simulations. The maximum absolute $\Delta_r$ and $\delta_h$values of Li–Li, Li–Si and Si–Si RDFs were used for each composition, which are referred as $|\Delta|_r^{max}$ and $|\delta|_h^{max}$, respectively.

## 3. Result and discussion

### 3.1 Structural similarity analysis

The similarity diagram in **Figure 2a** shows that configurations of $Li_1Si_{64}$ or $Li_{54}Si_1$ clustered into two small regions, respectively. This clustering suggests the existence of significant redundancy in configurations of these two compositions. Configurations of other three compositions in the training data set are relatively more disperse when compared with the clusters of $Li_{54}Si_1$ or $Li_1Si_{64}$. This trend indicates that the configurations in these three compositions have relatively low redundancy. We calculated the cumulative probability of pairwise similarity between configurations to quantitively compare the redundancy in different composition groups. The results in **Figure 2b** shows that a big portion of configurations in $Li_1Si_{64}$ or $Li_{54}Si_1$ composition group have a pairwise similarity larger than 0.95, which confirms that high redundancy exists in these two groups. Therefore, the size of the training data set can be significantly reduced if the redundancy is alleviated with a proper approach.

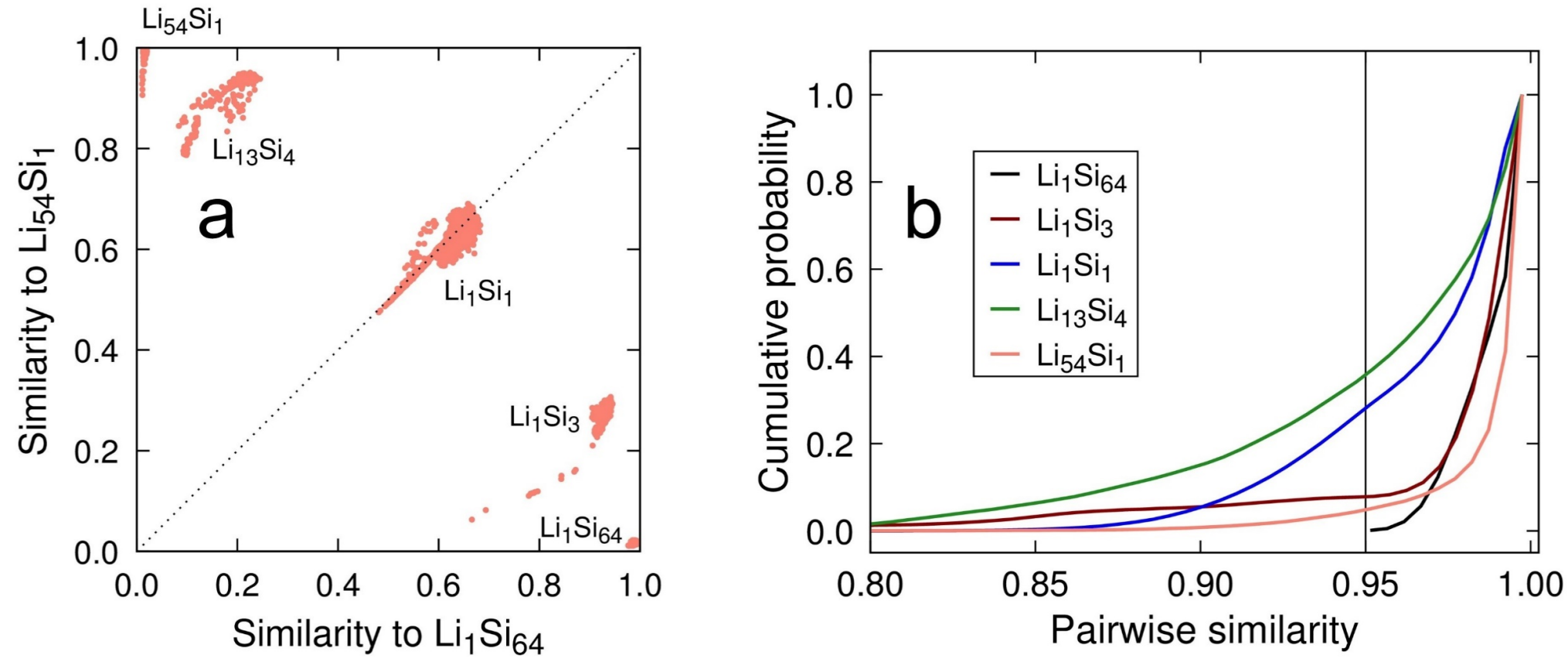

**Figure 2.** (a) Distributions of configurations from the training data set in a similarity diagram. (b) Cumulative probability of pairwise similarities between configurations with a specific composition in the training data set.

We then investigated the method to generate a refined data set which still retains the structural features of 3092 configurations in the training data set. A series of subsets ranging from 25 to 1500 configurations were constructed by using a combination of the structural similarity algorithm and the FPS method. The red line in **Figure 3** shows that the farthest distance between a configuration in the training data set and all configurations in an FPS subset are getting smaller with the increase of the data size, and converges to 0.1 when the data size reaches 200. A distance of 0.1 corresponds to a large similarity of 0.995. This trend indicates that when increasing the data size of an FPS subset containing more than 200 configurations, the new added configurations are quite similar to the existing configurations in the FPS subset. The blue line in **Figure 3** shows that the REP of an FPS subset is getting larger with the increase of the data size, and converges to 100% when the data size reaches 200. The fast converge of the REP confirms that the structural features in a small FPS subset such as containing 200 configurations may be as sufficient as those in the training data set.

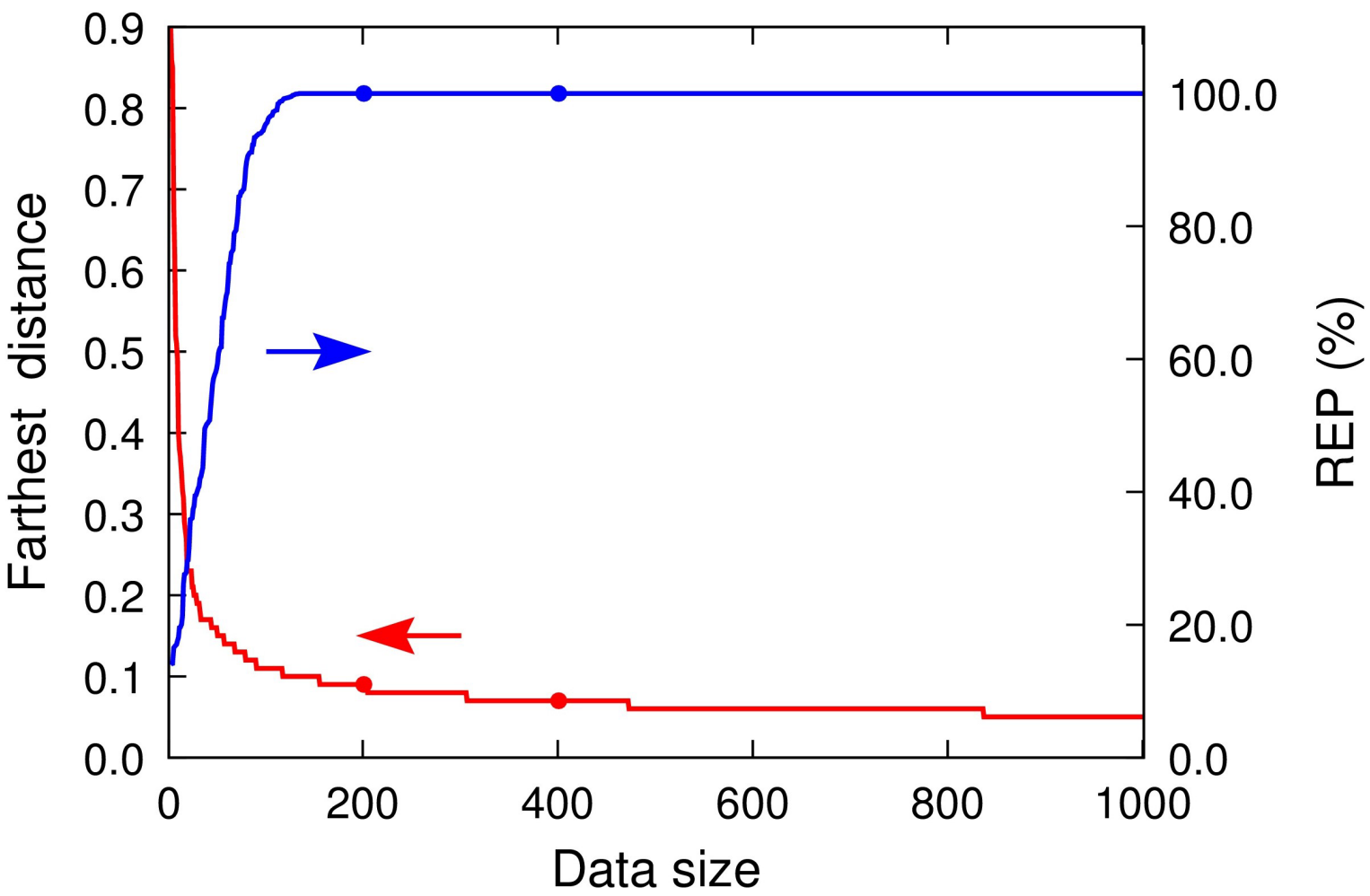

**Figure 3.** The farthest distance between a configuration in the training data set and all configurations in an FPS subset (red line) and the corresponding structural representativeness (REP) of the FPS subset (blue line) as a function of the data size. The FPS subsets of 200 and 400 configurations used for training MTPs are marked as dots in this figure.

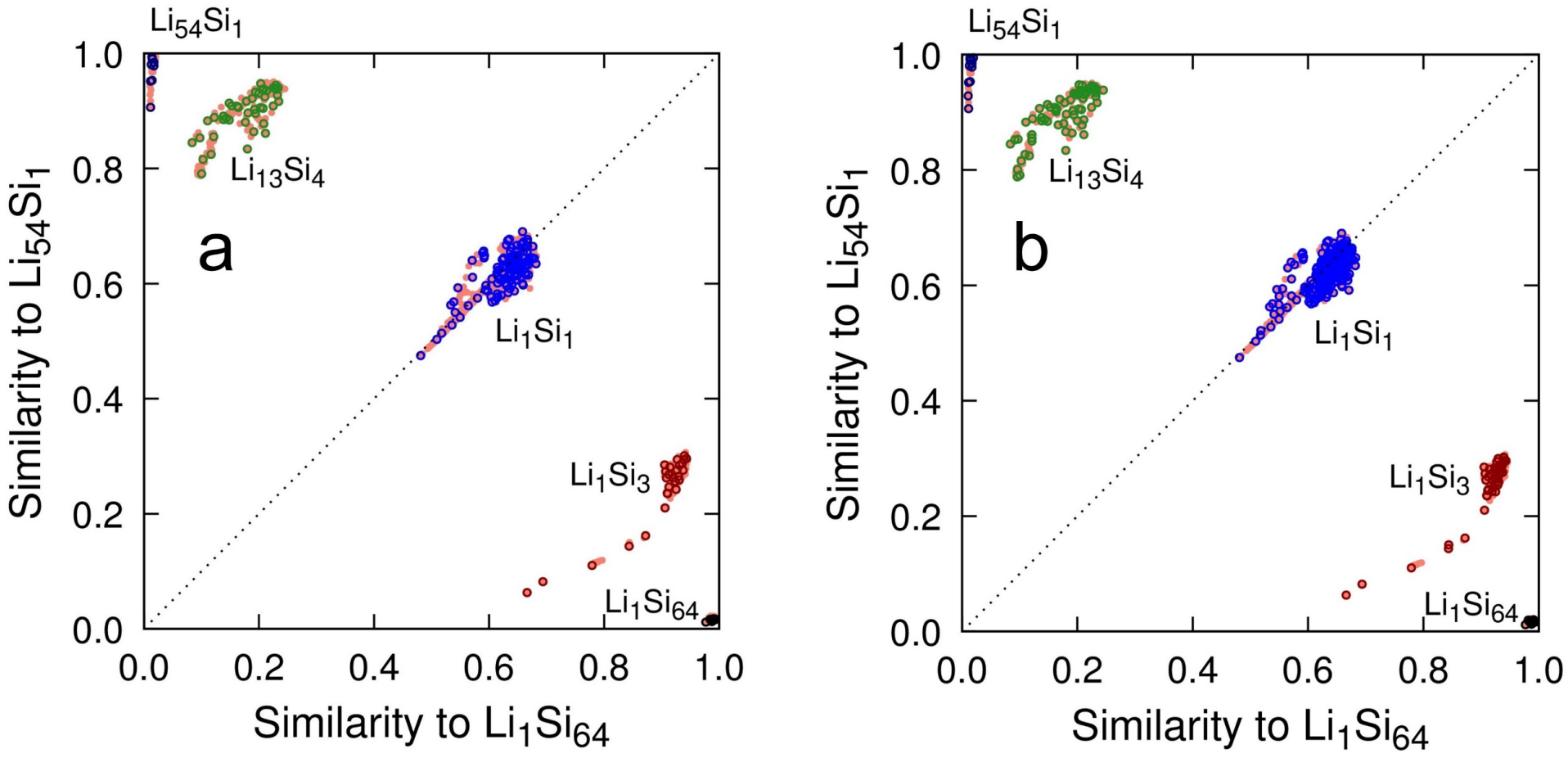

**Figure 4.** Distributions of configurations from (a) the FPS subsets containing 200 configurations, and (b) the FPS subset containing 400 configurations in similarity diagrams. The orange dots represent configurations from the training data set, while the black, red, blue, green and navy-blue dots represent configurations of $Li_1Si_{64}$, $Li_1Si_3$, $Li_1Si_1$, $Li_{13}Si_4$ and $Li_{54}Si_1$ in the FPS subsets, respectively.

Configurations from the FPS subsets containing 200 and 400 configurations were highlighted in similarity diagrams to visualize the coverage of these FPS subsets. As shown in **Figure 4a**, the data from the FPS subsets containing 200 configurations have covered most of data from the training data set, which also confirms that this FPS subset have sufficient representative of the training data set. It is noticeable that the FPS subset of 200 configurations only have few data for $Li_1Si_{64}$ and $Li_{54}Si_1$, which is also seen in **Table S1**. The decrease in percentages of configurations in $Li_1Si_{64}$ or $Li_{54}Si_1$ composition groups shows that the redundancy is remarkably reduced in the FPS subset. The FPS subset containing 400 configurations also has a sufficient coverage to the training data set in the similarity diagram, as shown in **Figure 4b**. In addition, it has a better coverage in some regions for $Li_{54}Si_1$, $Li_{13}Si_4$ and $Li_1Si_1$ than the FPS subset containing 200 configurations.

### 3.2 Potential energy and atomic force deviations

The similarity analysis shows that the FPS subsets containing 200 and 400 configurations have similar structural features as the training data set; nevertheless, we still need to evaluate the performance of the MTPs trained with these two FPS subsets by comparing the predicted potential energies, atomic forces and RDFs with those calculated by DFT and those predicted by the MTPs developed with the initial data set or the training data set.

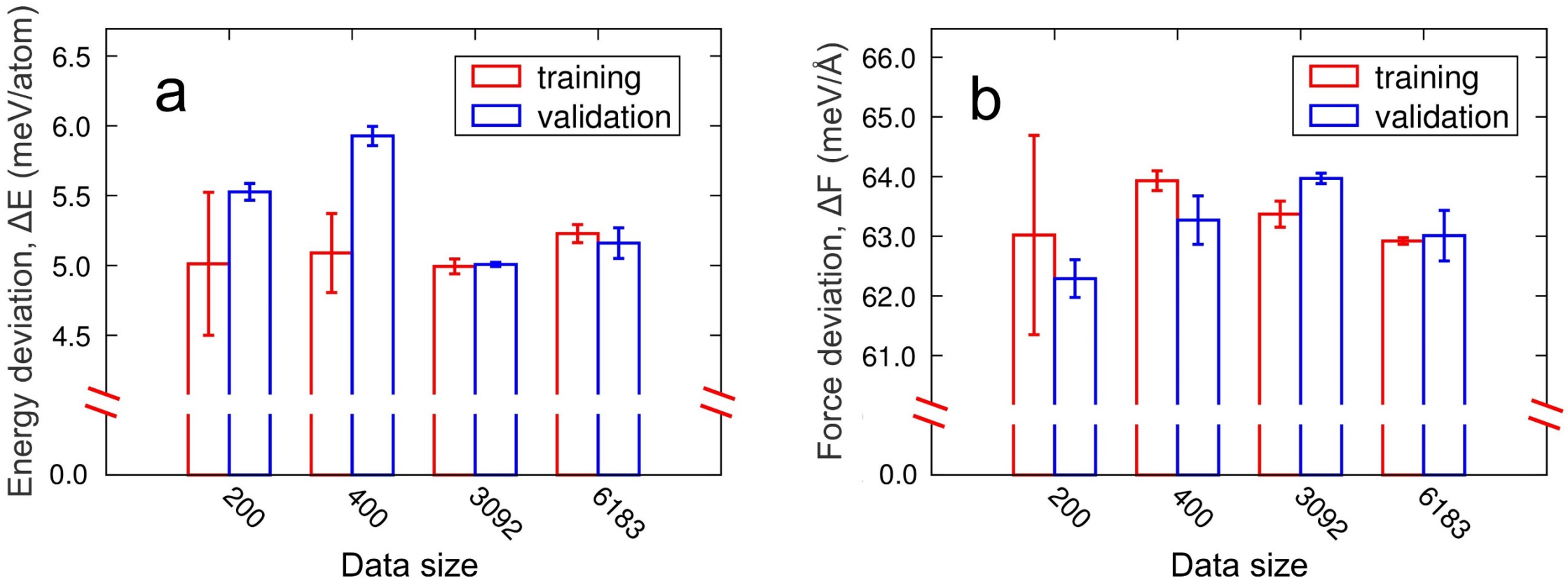

**Figure 5.** (a) The potential energy deviation ($\Delta E$) and (b) atomic force deviation ($\Delta F$) for training and validation predicted by the MTPs trained with the FPS subsets containing 200 and 400 configurations, the training data set, and the initial data set. The error bars indicate the standard errors for three batches with one batch contains about a third of configurations in the data set.

**Figure 5** shows that the MTPs trained with the FPS subsets containing 200 and 400 configurations can predict the potential energy and atomic forces for training and validation with an accuracy comparable to the MTPs trained with the training data set of 3092 configurations and the initial data set of 6183 configurations. The potential energy deviations for training and validation predicted by these four MTPs are in the range of 5.0 to 6.0 $meV/\text{atom}$, while the atomic force deviations for training and validation are in the range of 62.0 to 64.0 $meV/\text{Å}$. The potential energy deviations and atomic force deviations are also similar to other MTPs in the literature.[40, 51] The potential energy deviations predicted by the MTP developed for cathode coating materials[51] for training and validation are 4.5 and 5.7 $meV/\text{atom}$, while the atomic force deviations are about 94.7 and 84.3 $meV/\text{Å}$, respectively. It is noticeable that the MTP trained with the subset containing 200 configurations have significant fluctuations when predicting $\Delta E$ and $\Delta F$ for training, which indicates that the potential energies, atomic forces of configurations used for training differ remarkably and may limit the prediction ability of the MTP.[52]

The accuracy of the MTPs training with the FPS subsets containing 200 and 400 configurations in predicting atomic forces are further investigated by comparing the predicted values for the testing data set with those calculated by DFT. As shown in **Figure S1 and Figure S2**, the atomic forces on Li and Si atoms in amorphous Li−Si systems predicted by these two MTP and those calculated by the DFT distribute around the $y = x$ line narrowly. The narrow distributions indicate that the MTP trained with the FPS subsets containing 200 and 400 configurations achieve a reasonable accuracy in predicting atomic forces in the testing data set. This trend is confirmed by coefficient of determination $R^2$ for atomic forces on atoms in amorphous Si, $Li_1Si_1$, $Li_2Si_1$ and $Li_7Si_2$ shown in **Figure 6**. The $R^2$ for atomic forces on Li atoms by the MTPs trained with the FPS subsets containing 200 configurations and 400 configurations are in the range of 0.82 to 0.87, while the $R^2$ for atomic forces on Si atoms are in the range of 0.58 to 0.73, respectively. Most of the $R^2$ are higher than the corresponding values in different composition groups predicted by the MTP trained with the initial data set.

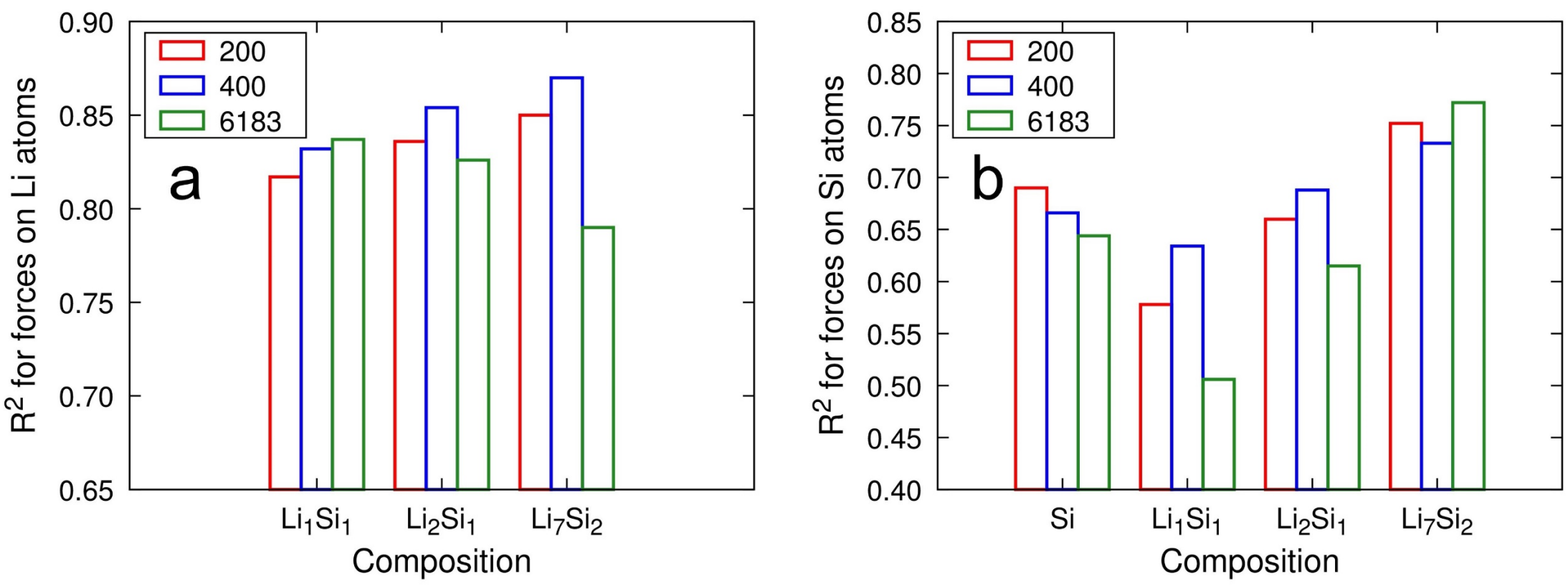

**Figure 6.** The $R^2$ for the atomic forces on (a) Li and (b) Si atoms calculated by DFT and predicted by MTPs trained with the FPS subsets containing 200 and 400 configurations, and the initial data set, respectively.

The higher $R^2$ for atomic forces on Li atoms than those for forces on Si atoms may originate from the fact that the energy of the Si−Si bond is more than twice that of the Li−Li bond.[53] Since the coordination of the Si−Si bonds is distorted in amorphous systems,[54] great forces exist on Si atoms. Therefore, large errors may accompany when predicting the forces on Si atoms. Interestingly, the $R^2$ for atomic forces on Si atoms gets larger with the increase of the Li−Si ratios. For example, the $R^2$ for forces on Si atoms in $Li_7Si_2$ and $Li_1Si_1$ by the MTP trained with the FPS subset containing 400 configurations are 0.73 and 0.63, respectively. The low concentration of Si in $Li_7Si_2$ decreases the possibility of forming Si−Si bonds, and therefore contribute to a higher $R^2$ for forces on Si atoms than that for forces on Si atoms in $Li_1Si_1$.

### 3.3 Predictions of RDFs

The potential energy and atomic forces deviation results show that the MTPs trained with the FPS subsets containing 200 and 400 configurations can predict the potential energies and atomic forces with an accuracy comparable to the MTP trained with the initial data set of 6183 configurations. To evaluate the accuracy of these MTPs in predicting structural properties, the Si−Si, Li−Si and Li−Li RDFs of amorphous Si, $Li_1Si_1$, $Li_2Si_1$ and $Li_7Si_2$ were calculated by MD simulations using these three MTPs, and compared with those calculated by the AIMD simulations.

The results in **Figure 7** shows that the MTP trained with the FPS subset containing 400 configurations predicts the locations and heights of the first RDF peaks in the Si−Si, Li−Si and Li−Li RDFs in $Li_7Si_2$ with an accuracy comparable to the AIMD simulation. The maximum deviations of the locations and heights of the first RDF peaks for other compositions in **Table 1** also confirm this trend. The maximum absolute position deviation $|\Delta|_r^{max}$ of the first RDF peaks for Si, $Li_1Si_1$, $Li_2Si_1$ and $Li_7Si_2$ calculated by this MTP are 0.00, 0.02, 0.04, and 0.04 Å, while the relative height deviation $|\delta|_h^{max}$ are 0.12, 0.12, 0.18 and 0.20, respectively. These small deviations in predicting the positions and heights of the first RDF peaks show that the MTP trained with the FPS subset containing 400 configurations successfully predicts the structural properties of amorphous Li−Si systems.

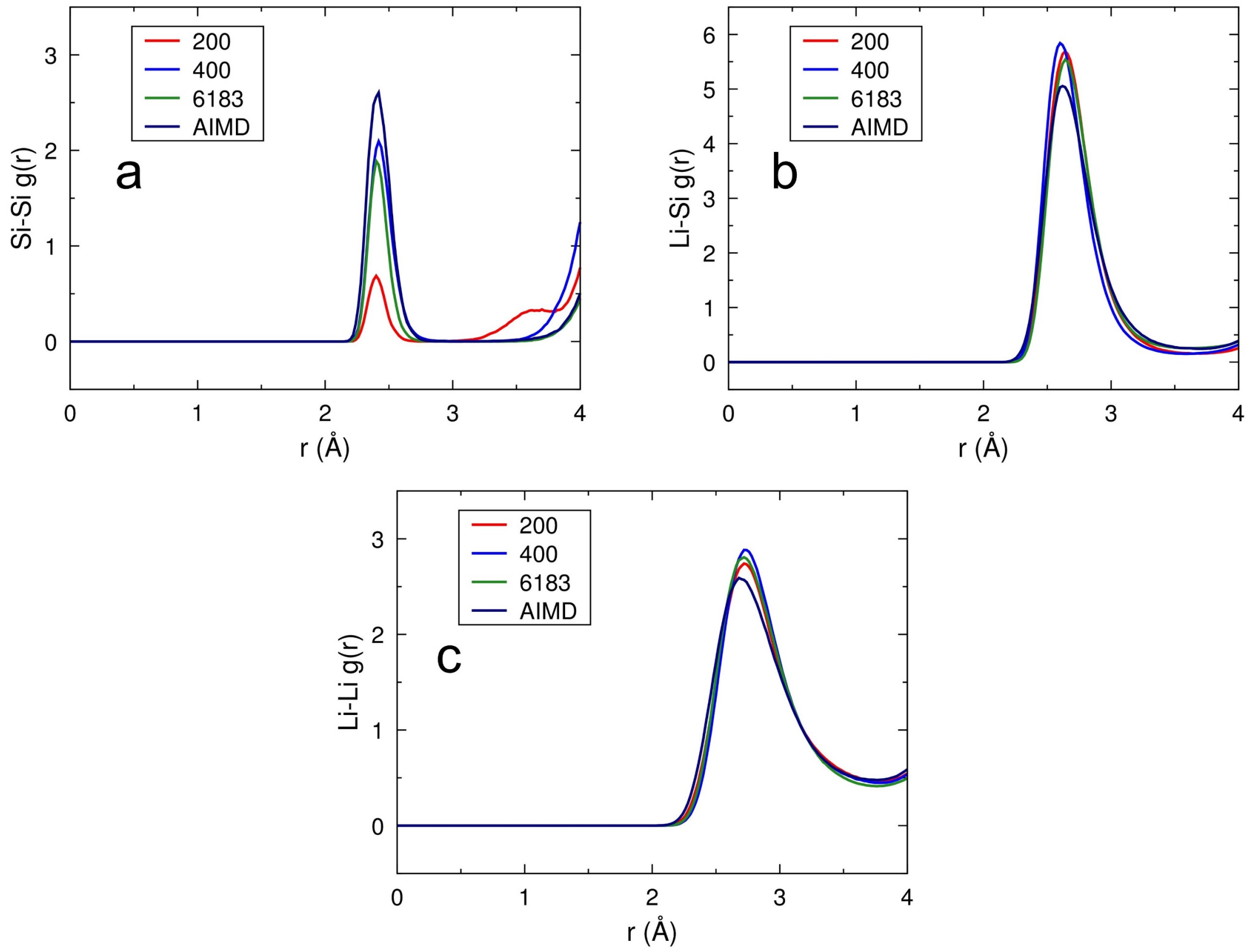

**Figure 7.** (a) Si–Si RDFs, (b) Li–Si RDFs and (c) Li–Li RDFs of amorphous $Li_7Si_2$ at 298 K from MD simulations using the MTPs trained with the FPS subsets containing 200 and 400 configurations, the initial data set of 6183 configurations compared with those from the AIMD simulation.

**Table 1. Maximum absolute position deviation $|\Delta|_r^{max}$ and relative height deviation $|\delta|_h^{max}$ of the first RDF peaks for Si, $Li_1Si_1$, $Li_2Si_1$, and $Li_7Si_2$**

| data size | $|\Delta|_r^{max}$ (Å)/$|\delta|_h^{max}$ | | | |
|---|---|---|---|---|
| | Si | $Li_1Si_1$ | $Li_2Si_1$ | $Li_7Si_2$ |
| 200 | 0.00/0.09 | 0.04/0.30 | 0.04/0.17 | 0.04/0.74 |
| 400 | 0.00/0.12 | 0.02/0.12 | 0.04/0.18 | 0.04/0.20 |
| 6183 | 0.00/0.09 | 0.04/0.18 | 0.04/0.17 | 0.04/0.18 |

**Figure 7** also demonstrates that the FPS subset containing 200 configurations have large deviations in predicting the Si−Si RDF in $Li_7Si_2$. It predicts a much lower peak of Si−Si RDF in $Li_7Si_2$ compared to that by the AIMD simulation, which is also shown in **Table S2**. The molar fraction of Si in $Li_7Si_2$ is as low as 0.22, therefore, Li−Si pairs instead of Si−Si pairs would be dominating for each Si atom, which will greatly increase the difficulty of sampling Si−Si interactions. The MTP trained with the FPS subset containing 400 configurations has a relatively more configurations in $Li_{13}Si_4$ composition group with a molar fraction of Si similar to $Li_7Si_2$ when compared to the FPS subset of 200 configurations. Therefore, it better predicts the location and height of the first Si−Si RDF peak.

### 3.4 Comparisons between the FPS and classic SS methods

We also investigated the performance of MTPs trained with 3 subsets containing 400 configurations selected by the stochastic sampling. **Figure 8** shows that the $R^2$ for forces on Si atoms in $Li_1Si_1$, $Li_2Si_1$, $Li_7Si_2$ by these three MTPs are remarkably lower than that calculated by the MTP trained with the FPS subset containing 400 configurations, as shown in **Table S3**. Moreover, the $R^2$ for forces on Si atoms predicted by the MTPs training with the SS subsets have large fluctuations, which may arise from the randomness induced by the stochastic sampling. As shown in **Table S4**, the percentages of compositions in the SS subsets were maintained almost the same as those in the initial data set. However, some essential and representative configurations in the initial data set may be missing when using the stochastic sampling. As the FPS method constructs subsets by selecting representative configurations based on the structural similarity analysis, it has the advantage over the SS method in retaining representative configurations when refining the training data set.

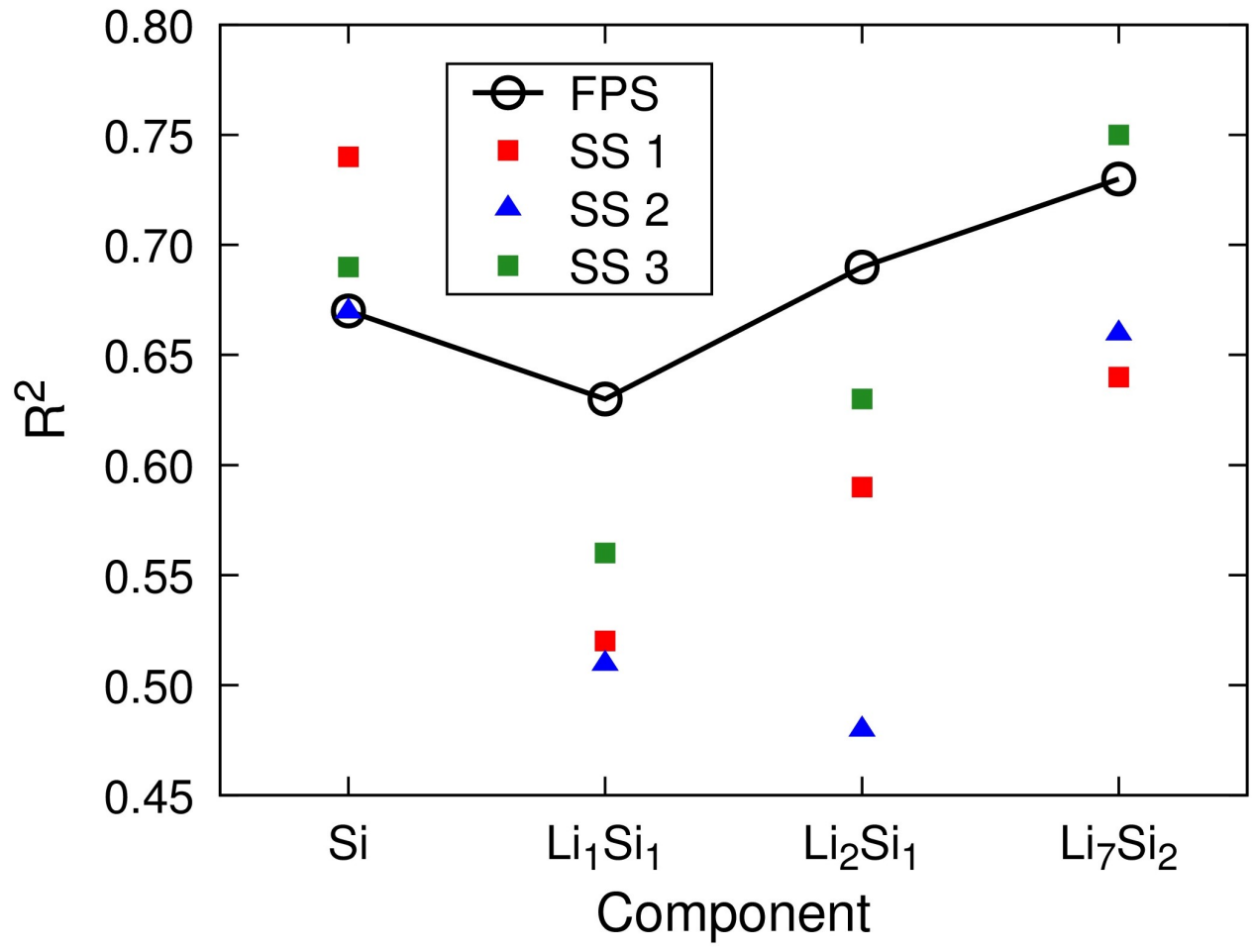

**Figure 8.** The $R^2$ values for forces on Si atoms in the testing set predicted by the MTPs trained with the FPS subset containing 400 configurations and three SS subsets containing 400 configurations.

## 4. Conclusion

This work investigates the method to reduce redundancy from an initial training data set without losing the quality of the developed machine learning potential using a combination of a structural similarity algorithm and the FPS method. We calculated the structural representativeness of the FPS subsets with different data sizes, and found that an FPS subset containing more than 200 configurations has the ability to represent the training data set containing 3092 configurations. According to the similarity analysis, we trained MTPs using the FPS subsets containing 200 and 400 configurations, and then evaluated the performance of these two MTPs on predicting the potential energies, atomic forces and RDFs of Li–Si systems. The simulation results show that the MTP trained with the FPS subset containing 400 configurations can achieve an accuracy comparable to the one developed based on the initial data set of 6183 configurations. We also found that the similarity-based FPS method has the advantage over the classic stochastic sampling method in retaining representative configurations when refining the training data set. Our work highlights that the similarity-based FPS method could reduce redundancy in different compositions, and therefore accelerate the development of a high-accurate machine learning potential.

## AUTHOR INFORMATION

Corresponding Author

*E-mail: yihezj@zju.edu.cn

## NOTES

The authors declare that there is no conflict of interest.

## ACKNOWLEDGEMENT

N.X., Y.S., and Y.H. express thanks for the financial support from the National Key Research and Development Program of China (grant number 2018YFC0213806) and the Leading Innovative and Entrepreneur Team Introduction Program of Zhejiang (grant number 2019R01006). Q.S. would like to thank the financial support provided by the Startup Funds of the University of Kentucky.

**FOR TABLE OF CONTENTS ONLY**

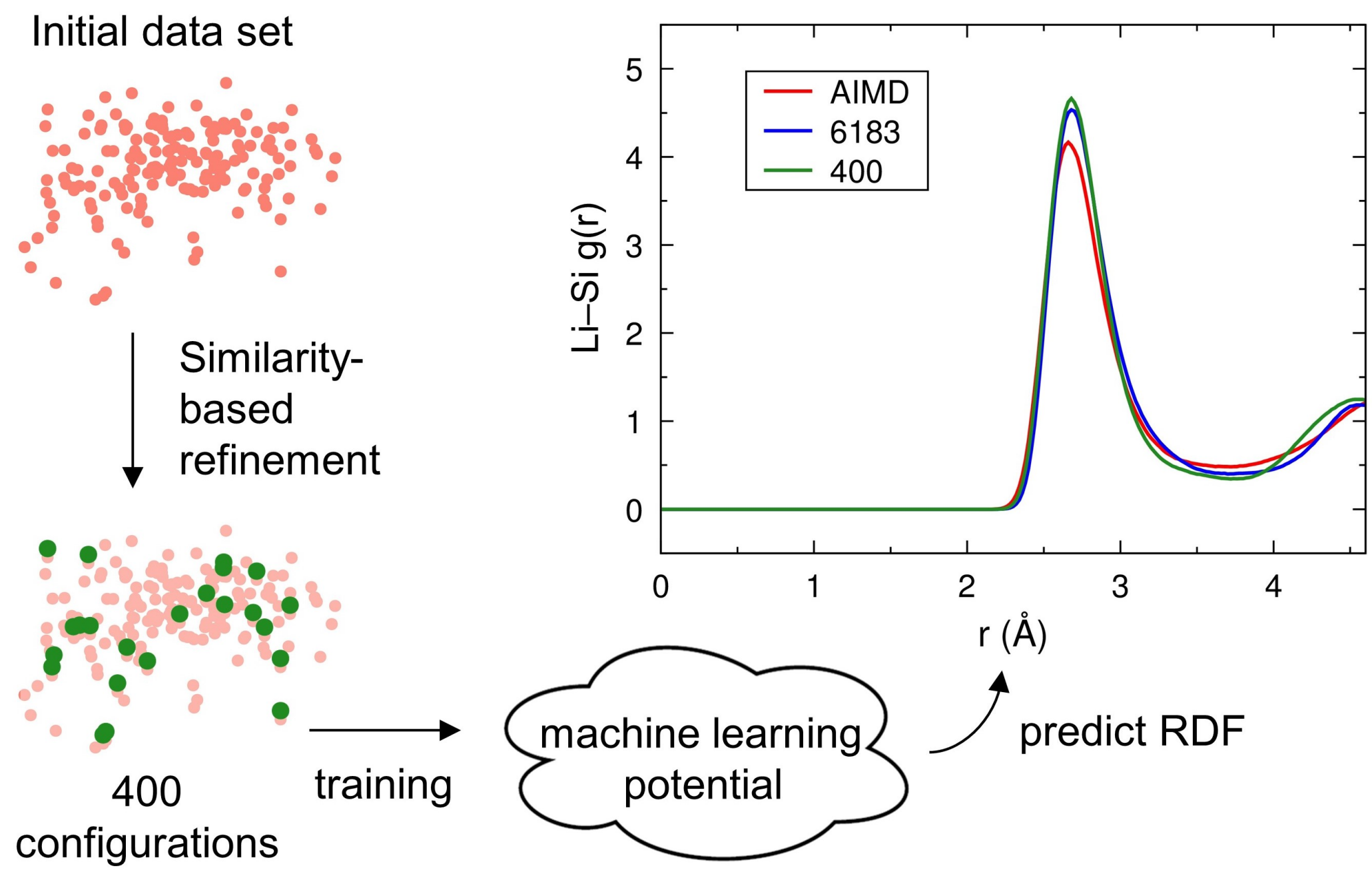

**Supplementary information for: "Training Data Set Refinement for the Machine Learning Potential of Li–Si Alloys via Structural Similarity Analysis"**

*Nan Xu,*[a] *Chen Li,*[a] *Mandi Fang,*[a] *Qing Shao,*[b] *Yao Shi,*[a] *Yingying Lu,*[a] *Yi He,*[a,c*]

a College of Chemical and Biological Engineering, Institute of Zhejiang University-Quzhou, Key Laboratory of Biomass Chemical Engineering of Ministry of Education, Zhejiang University, Hangzhou 310027, China

b Chemical and Materials Engineering Department, University of Kentucky, Lexington, KY 40506, USA

c Department of Chemical Engineering, University of Washington, Seattle, WA 98195, USA

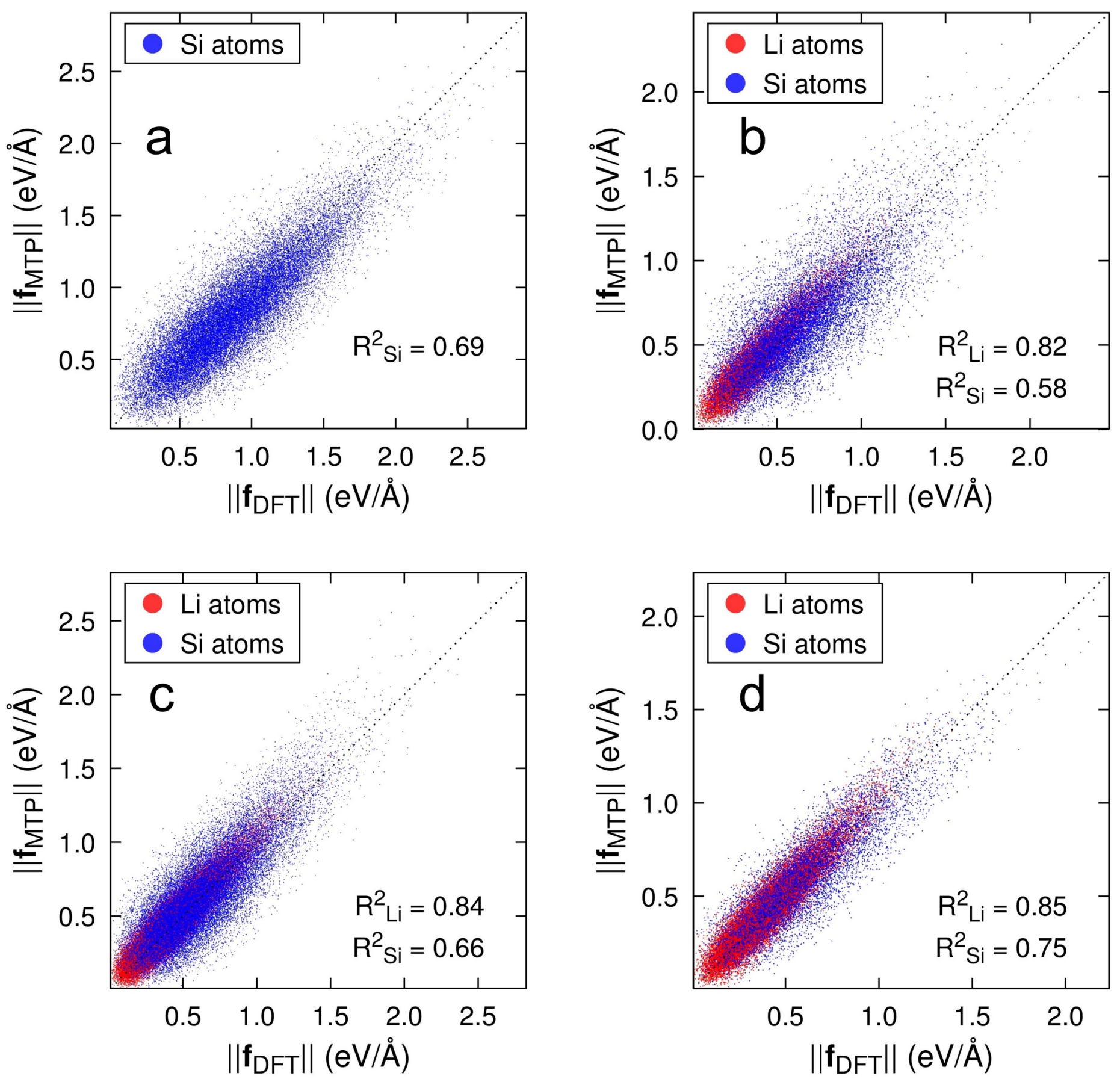

**Figure S1.** Atomic forces on Si and/or Li atoms predicted by the MTP trained with the FPS subset containing 200 configurations compared with the DFT results for amorphous (a) Si, (b) $Li_1Si_1$, (c) $Li_2Si_1$, and (d) $Li_7Si_2$ in the testing data set.

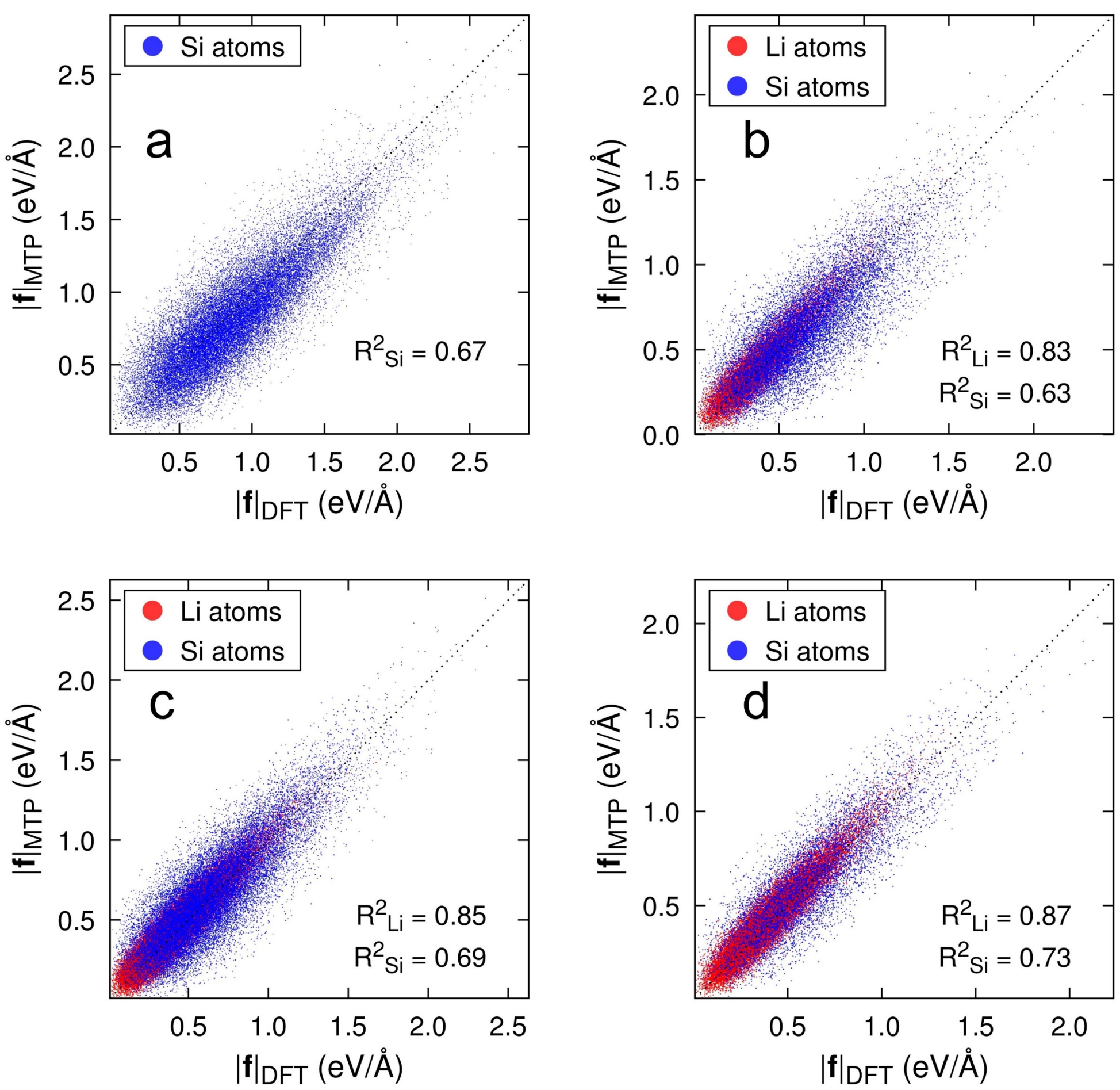

**Figure S2.** Atomic forces on Si and/or Li atoms predicted by the MTP trained with the FPS subset containing 400 configurations compared with the DFT results for amorphous (a) Si, (b) $Li_1Si_1$, (c) $Li_2Si_1$, and (d) $Li_7Si_2$ in the testing data set.

**Table S1. Numbers and percentages of configurations for each composition in a specific FPS subset, the training data set, the validation set, or the initial data set**

| data size | Numbers of configurations/Percentages of configuration (%) | | | | |
|---|---|---|---|---|---|
| | $Li_1Si_{64}$ | $Li_1Si_3$ | $Li_1Si_1$ | $Li_{13}Si_4$ | $Li_{54}Si_1$ |
| 25 | 2/8.00 | 5/20.00 | 7/28.00 | 7/28.00 | 4/16.00 |
| 50 | 4/8.00 | 9/18.00 | 19/38.00 | 13/26.00 | 5/10.00 |
| 100 | 5/5.00 | 16/16.00 | 45/45.00 | 25/25.00 | 9/9.00 |
| 150 | 7/4.67 | 23/15.33 | 76/50.67 | 35/23.33 | 9/6.00 |
| 200 | 9/4.50 | 27/13.50 | 112/56.00 | 43/21.50 | 9/4.50 |
| 400 | 11/2.75 | 48/12.00 | 257/64.25 | 71/17.75 | 13/3.25 |
| 600 | 13/2.17 | 75/12.50 | 407/67.83 | 90/15.00 | 15/2.50 |
| 800 | 16/2.00 | 94/11.75 | 567/70.88 | 104/13.00 | 19/2.38 |
| 1000 | 18/1.80 | 109/10.90 | 730/73.00 | 124/12.40 | 19/1.90 |
| 1500 | 21/1.40 | 152/10.13 | 1141/76.07 | 164/10.93 | 22/1.47 |
| 3092[a] | 290/9.38 | 286/9.25 | 1963/63.49 | 290/9.38 | 263/8.51 |
| 3091[b] | 290/9.38 | 286/9.25 | 1963/63.51 | 290/9.38 | 262/8.48 |
| 6183[c] | 580/9.38 | 572/9.25 | 3926/63.50 | 580/9.38 | 525/8.49 |

a, b, c denotes the training data set, the validation data set and the initial data set.

**Table S2. Position deviation $\Delta_r$ and relative height deviation $\delta_h$ of the first RDF peaks in $Li_7Si_2$**

| data size | $\Delta_r$(Å) | | | $\delta_h$ | | |
|---|---|---|---|---|---|---|
| | Si−Si | Li−Si | Li−Li | Si−Si | Li−Si | Li−Li |
| 200 | 0.02 | 0.02 | 0.04 | -0.74 | 0.13 | 0.06 |
| 400 | 0.00 | -0.02 | 0.04 | -0.20 | 0.16 | 0.11 |
| 6183 | -0.02 | 0.02 | 0.04 | -0.18 | 0.10 | 0.05 |

**Table S3. $R^2$ for forces on Li and Si atoms in the testing data set by the MTP trained with the initial data set, and MTPs trained with the FPS subset containing 400 configurations and three SS subsets containing 400 configurations**

| data size | $R^2$ of atomic force errors for Li/Si atoms | | | |
|---|---|---|---|---|
| | Si | $Li_1Si_1$ | $Li_2Si_1$ | $Li_7Si_2$ |
| 6183 | -/0.64 | 0.84/0.51 | 0.83/0.62 | 0.79/0.77 |
| $400^{*}$ | -/0.67 | 0.83/0.63 | 0.85/0.69 | 0.87/0.73 |
| $400^{\#1}$ | -/0.74 | 0.82/0.52 | 0.84/0.59 | 0.83/0.64 |
| $400^{\#2}$ | -/0.67 | 0.86/0.51 | 0.85/0.48 | 0.85/0.66 |
| $400^{\#3}$ | -/0.69 | 0.82/0.56 | 0.82/0.63 | 0.83/0.75 |

- means no data here, * denotes the FPS subset and # stands for three SS subsets.

**Table S4. Numbers and percentages of configurations for each composition in the initial data, the FPS subset of 400 configurations and three SS subsets of 400 configurations**

| data size | sample numbers/percentages (%) of each component | | | | |
|---|---|---|---|---|---|
| | $Li_1Si_{64}$ | $Li_1Si_3$ | $Li_1Si_1$ | $Li_{13}Si_4$ | $Li_{54}Si_1$ |
| 6183 | 580/9.38 | 572/9.25 | 3926/63.50 | 580/9.38 | 525/8.49 |
| $400^*$ | 11/2.75 | 48/12.00 | 257/64.25 | 71/17.75 | 13/3.25 |
| $400^{\#1}$ | 28/7.00 | 40/10.00 | 271/67.75 | 31/7.75 | 30/7.50 |
| $400^{\#2}$ | 31/7.75 | 30/7.50 | 277/69.25 | 40/10.00 | 22/5.50 |
| $400^{\#3}$ | 33/8.25 | 38/9.50 | 252/63.00 | 33/8.25 | 44/11.00 |

* denotes the FPS subset and # stands for three SS subsets.